# Genetic variation in human drug-related genes.

Charlotta P.I. Schärfe[1,2,3], Roman Tremmel[4], Matthias Schwab[4,5,6], Oliver Kohlbacher[2,3,7,8,9*], Debora S. Marks[1*]

[1] Department of Systems Biology, Harvard Medical School, Boston, 02115 Massachusetts, USA

[2] Center for Bioinformatics, University of Tübingen, 72076 Tübingen, Germany

[3] Applied Bioinformatics, Dept. of Computer Science, 72076 Tübingen, Germany

[4] Dr. Margarete Fischer-Bosch Institute of Clinical Pharmacology, Stuttgart, Germany

[5] Department of Clinical Pharmacology, University Hospital Tübingen, Germany,

[6] Department of Pharmacy and Biochemistry, University of Tübingen, Tübingen, Germany

[7] Quantitative Biology Center, 72076 Tübingen, Germany

[8] Faculty of Medicine, University of Tübingen, 72076 Tübingen, Germany

[9] Biomolecular Interactions, Max Planck Institute for Developmental Biology, 72076 Tübingen, Germany

* Corresponding authors:

E-mail: oliver.kohlbacher@uni-tuebingen.de, debbie@hms.harvard.edu

# Abstract

Variability in drug efficacy and adverse effects are observed in clinical practice. While the extent of genetic variability in classical pharmacokinetic genes is rather well understood, the role of genetic variation in drug targets is typically less studied. Based on 60,706 human exomes from the ExAC dataset, we performed an in-depth computational analysis of the prevalence of functional-variants in in 806 drug-related genes, including 628 known drug targets. We find that most genetic variants in these genes are very rare ($f < 0.1\%$) and thus likely not observed in clinical trials. Overall, however, four in five patients are likely to carry a functional-variant in a target for commonly prescribed drugs and many of these might alter drug efficacy. We further computed the likelihood of 1,236 FDA approved drugs to be affected by functional-variants in their targets and show that the patient-risk varies for many drugs with respect to geographic ancestry. A focused analysis of oncological drug targets indicates that the probability of a patient carrying germline variants in oncological drug targets is with 44% high enough to suggest that not only somatic alterations, but also germline variants carried over into the tumor genome should be included in therapeutic decision-making.

About three in five Americans aged 20 and above take prescription drugs every month[1] and many either encounter adverse drug reactions or reduced treatment efficacy[2]. The strong genetic component of altered drug response in patients is well known[3] and attributed to variants affecting drug pharmacokinetics (PK) and pharmacodynamics (PD)[4]. Methods to identify these genetic determinants have been developed in population stratified[5-7] or individualized settings[4,8]. Particularly, the vast amount of genetic information now available has opened up the possibility to systematically study inter-individual differences in drug response using genome-wide association (GWA) studies[9,10]. Results of these efforts have so far led to the pharmacogenomics labeling of 170 drugs by the Food and Drug Administration (FDA)[11] and the establishment of pharmacogenomics screening in many large hospitals in the US[12] and Europe[13].

However, typical pharmacogenomics GWA studies struggle with study sizes that are only large enough to detect common variants with an effect on the phenotype, but are unable to statistically pick up signals from rare variants with a functional effect[9,10]. Thus, data from recent genetic population catalogs such as the 1,000 Genomes project[14] and the NHLBI Exome Sequencing Project (ESP) have been used to determine the spectrum of variation in pharmacokinetics-related genes. While classification of common and rare varies by study, especially variants considered to be on the rare end of the spectrum (minor allele frequency (minor AF) < 0.5%) were found abundantly in genes associated with drug absorption, distribution, metabolism, or excretion (ADME)[15,16] as well as in potential drug targets[17]. Based on these surveys, it was estimated that at least 97% of individuals carry actionable high-risk pharmacological variants affecting drug ADME in their genome[12,18]. However, the role of genetic variation in pharmacologically established drug targets is less well studied.

The Exome Aggregation Consortium (ExAC)[19] has aggregated data from several large sequencing studies comprising exome sequencing data of 60,706 individuals – nearly an order of magnitude larger than the public population catalogs mentioned above. Using a cohort of this size, it now becomes possible to study even very rare variants in drug target and ADME genes and to calculate the overall risk of containing a functional-variation for each patient. Furthermore, even though geographic ancestry is a known confounding factor for drug response and has been incorporated in clinical decision making in the absence of individual genotype data[20], a comprehensive inventory of functional genetic variation in drug-associated genes across populations is still lacking. A cohort of the size of the ExAC catalog now allows determining the allele frequency of very rare variants in distinct population sub-groups and comparing their prevalence.

In this study, we provide a comprehensive analysis of genetic variation predicted to result in altered protein function ("functional-variants") in 806 drug-related genes including 628 drug targets (163 targeted by cancer-therapeutics). We further describe how this may affect the likelihood of 1,236 FDA approved drugs to be affected by functional-variants in their targets and how this likelihood varies between different populations.

## Results

**Drug-related genes show high extent of genetic variability across 60K individuals**

To explore the extent of non-synonymous genetic variation in drug-related genes in the human populations, we analyzed single nucleotide variants in 60,706 human individual exomes from ExAC[19] in a set of 806 drug-related genes collated from DrugBank[21] and other sources[15,22] (Fig. 1a, Supplementary Table 1). The AF distribution of non-synonymous variants in drug-related genes is almost identical to that of all genes (n=17,758) and 97.5% of observed non-synonymous variants have an allele frequency < 0.1% (sometimes termed a "rare variant"[19]) (Fig. 1b, Supplementary Fig. 1). Of note, 71% of the variants in the human exome, including drug-related genes have not been observed previously in public repositories such as dbSNP and therefore can be considered novel (Supplementary Fig. 1).

To identify variants that are most likely to affect the gene function ("functional-variants"), we filtered the set of non-synonymous variants for those resulting in the loss of the protein product ("loss-of-function", LoF)[19], or predicted to be "damaging" by PolyPhen-2[23] and SIFT[24]. This resulted in 61,134 functional-variants in 806 drug-related genes (of which 767 genes included at least one LoF variant) and, not surprisingly, these functional-variants tend to have lower AFs than all other non-synonymous variants (98.7% have an allele frequency < 0.1%) (Fig. 1c). Nevertheless, 43% of the drug-related genes with predicted functional-variants have at least one functional-variant with AF ≥ 0.1%. The drug-related genes with the most frequent functional-variants are membrane transporter genes related to drug efflux and uptake such as *ABCB5* (three LoF, six damaging), *SLC22A1* (nine damaging), and *SLC22A14* (eight damaging). In the clinically highly important polymorphic cytochrome P450 enzyme *CYP2D6* also eight damaging variants have been identified (Supplementary Table 2). Since the ExAC cohort contains an order of magnitude more individuals than previously available, it also allowed us to identify genes with many different functional-variants even though each variant may be individually rare. The ADME genes with the most functional-variants per residue reflect similar findings from smaller cohort studies and include the glutathione S-transferase sodium/bile transporter *SLC10A1* (0.36 variants/residue), *GSTA5* (0.31 variants/residue), and some cytochromes P450s such as *CYP1A1* (0.30 variants/residue) and *CYP2C19* (0.28 variants/residue)[15]. Furthermore, our analysis revealed drug target genes with comparable numbers of functional-variants per residue including the dofetilide target *KCNJ12* (0.31 variants/residue) and the target for the rheumatoid arthritis drug niflumic acid, *PLA2GLB* (0.30 variants/residue) (Supplementary Table 3).

While both metrics described above may be useful to evaluate the extent of genetic variation in the human population, they do not quantify the risk of an individual person in the population to carry functional-variants in a particular gene. In order to estimate this risk, we define a statistic, the "cumulative allele probability" (CAP), which captures both the number of functional-variants and their allele frequencies per gene (Methods and Supplementary Table 2). We want to emphasize that the CAP score of a gene does not necessarily reflect the extent to which the variants change the pharmacological behavior of the drug and therefore should be regarded as a score solely indicating a potential pharmacogenetic risk. Amongst the genes with the highest CAP scores, that is the highest probability of being affected by a functional-variant, are both, ADME genes and drug targets. The ADME genes with the highest CAP scores include *NAT2* (81%, involved in metabolizing arylamine and hydrazine drugs), *CYP2D6* (59.6%, involved in the metabolism of 20% of most prescribed drugs in the US[25]) and the transporter gene *SLCO1B1* (26.0%, a high risk gene for simvastatin-related myopathy/rhabdomyolsis[26]). The drug target genes with comparable high CAPs scores include tyrosinase (*TYR*; 62.4%, targeted by the acne drug azelaic acid), the alpha-4 subunit of the $GABA_A$ receptor *GABRA4* (53%, targeted by benzodiazepines) and *F5* (20.1%, targeted by drotrecogin alpha which was withdrawn from the market due to unacceptable high number of adverse drug reactions) (Fig. 2). The major proportion of the CAP score for these highest 'risk' genes derives from common genetic variants many of which have been observed previously. Nevertheless, for many genes a non-negligible proportion of the score is contributed by rare functional-variants, which were identified through the sufficiently large cohort size (see the lines in light purple and light blue in Figure 2a and 2b, respectively and Supplementary Table 2). In addition, we estimate that more than 60% of the drug-related genes in our set are putative novel candidates for pharmacogenomic research, so far missing relevant information from clinical studies (Supplementary Fig. 2)[27].

**Cancer drug target genes have many germline functional-variants.**

Especially in cancer therapy, genetic variation in drug targets has been recognized to play a crucial role for treatment success[28,29]. While some cancer drugs do not act in the tumor tissue, the cancer drug's primary site of action usually is in the tumor, whose genome contains tumor-specific somatic variants as well as a subset of patient-specific germline variants[30]. Information on somatic variants from tumor samples is thus increasingly used to enable research on drug design and to implement stratified or personalized cancer therapy. However, the patient's germline genome is routinely masked in these tumor sequencing analysis protocols[28,29] We thus wanted to assess whether target genes of drugs used in cancer therapy contain germline variants in the population that may affect the drug action and may be missed by current tumor sequencing analysis protocols. More than 15% of the drugs in this report (193 of the 1,236) are used in oncology (as defined by the WHO ATC code[31]) and between them have 163 gene targets. Several of these targets have high probabilities of having a functional-variant in the germline (Supplementary Table 2). For some of these targets the germline risk directly corresponds to potential altered treatment effects. This is the case for the kinase *KDR* (also known as *VEGFR2)* (CAP=25%), which is targeted by sorafenib and sunitib to inhibit vascularization of the tumor site[32]. Other drug targets for cancer therapeutics with high CAP scores include *MAP4* (60%) and *TUBB1* (30%) that are targets of paclitaxel, *MAP1A* (42%) a target of estramustine, *CD3G* (39%) a target of muromonab and *PARP1* (37%) a target of olaparib (Fig. 2). Overall, 40 cancer drug target genes, including 34 target genes with kinase domains, show CAP scores >1%. For these examples, functional germline variants are only relevant for treatment response if the tumor genome also carries them. While there is not a complete overlap between both germline and tumor genome due to loss of heterozygosity and other alterations in carcinogenesis[30], our analysis suggests that a large percentage of the population may contain functional-variants in cancer therapeutic targets in the germline that may carry over to the cancer genome and could be easily overlooked by current analysis protocols.

**Aggregating risk for functional-variants in targets by drug highlights drug candidates for future pharmacogenomics research**

About 70% of the FDA-approved drugs analyzed here do not have any pharmacogenomics data associated with them in public repositories[27]. However, our analysis shows that there are many functional-variants in their target genes (Fig. 3a). To estimate how much each drug can be affected by functional-variants in its target genes and to highlight possible candidates for future research, we computed the probability of containing a functional-variant in any number of its reported targets in DrugBank[21] by combining the CAP scores of the drug's target genes to a "drug risk probability" (short DRP, see Methods for details). For all FDA-approved drugs considered here (n=1,236), 43% have a DRP greater than 1% (Supplementary Table 4). The DRPs are weakly correlated to the number of targets (linear regression, $r^2 = 0.28$), leaving many drugs with few targets but higher than expected DRPs (determined by root mean square errors, short RMSE, of the model, red circles in Supplementary Fig. 3). For instance, one of the two human targets of azelaic acid, tyrosinase (*TYR*) is highly mutated in the population causing a DRP of 62.5% for this drug, which results in an RMSE of 0.34.

Drugs with the top DRP scores are paclitaxel and docetaxel (82%), quinacrine (70%), azelaic acid (63%), triazolam and other benzodiazepines (>50%) (Supplementary Table 4). This means that any individual in the population has a probability of more than 50% to carry a functional-variant that may affect the medication outcome of these drugs. Several of the drugs with high DRPs are considered "essential medicines" by the WHO[33]. In addition to paclitaxel and docetaxel, these include the opioid methadone (13.6%), the diuretic amiloride (11.7%), and the local anesthetic lidocaine (11.4%). For instance, the drug methadone targets the D- and M-type opioid receptors (*OPRD1, OPRM1*) and whilst some non-coding variants and a single coding variant (rs1799971) have previously been associated with required dose adjustments and treatment response, we observe another 132 functional-variants in these target genes, which could therefore be candidates for further testing. Since variants with predicted damaging effects dominate especially the rather high DRPs, we filtered the variants for only those resulting in LoF. Restricting to these high confidence variants, the DRP decreases below 10% and the drugs with the highest DRP include the anti-cancer drug marimastat (8.3%), the anti-ulcer medication sulfacrate (8.2%), the anti-flu drug oseltamivir (6.0%) which targets human *CES1* for activation, and several liptins used for diabetes that inhibit *DPP4* (5.6%) (Supplementary Table 4).

We then focused our analysis on the top 100 most prescribed medications in the US (from 2013[34]) which results in a list of 77 unique drug compounds for further investigation. 42% of these drugs have a DRP score greater than 1% of containing a functional-variant and the probability of an individual carrying a functional-variant in any of the targets for these 77 top prescribed drugs is 81%. For some of these drugs it is already well established that there is some genetic component to drug response, even if the details are debated[35]. For instance, five of the top fifteen most prescribed drugs in the US are asthma drugs (budesonide, salbutamol, salmeterol, fluticasone, and tiotropium). Whilst each of the DPRs is not particularly high (ranging from 0.06% to 0.25%), their widespread prescription rate (> 100 million prescriptions in 2013) still results in thousands of individuals who may be affected by a functional-variant. Similarly, statins (e.g., atorvastatin and rosuvastatin) are prescribed to nearly one in five adults in the US[1] and primarily target *HMGCR*. Due to genetic variation in this target gene statins have a DRP of 0.18%. This means that of the 40 million individuals who are prescribed a statin in the US, more than 80,000 individuals could be at risk of altered pharmacodynamics of statin treatment due to a functional-variant in the target *HMGCR*. This finding is underlined by previous pharmacogenetic studies showing that *HMGCR* is the most important polymorphic gene for treatment success of statins[36].

Overall, the genetic-variability of drug targets of many of the top 100 prescribed drugs has not been systematically annotated so far (Supplementary Fig. 4), including the Alzheimer's drug memantine (DRP=7.2%), the pain-medication acetaminophen (DRP=4.7%) and the proton-pump inhibitor esomeprazole (DRP=3.1%) that all have high DRPs. While these drugs, to our knowledge, do not have known associations between functional-variants in drug targets with drug action, clinical studies show that certain proportions of patients treated with them do not respond to treatment. The extent of this non-response is reflected by the number needed to treat, NNT[37]. For instance, for every one patient successfully treated for Alzheimer's diseases with memantine, between two and seven patients do not respond to treatment[38] (NNT=3 to 8). Similarly, the NNT for acetaminophen and its indication of pain is five[39] and for esomeprazole and reflux disease is 54[40].

**Drug-related genes show geographic difference in genetic variability.**

It is known that individuals with different geographic ancestry carry genetic variants with different frequencies[41]. The six populations differentiated in ExAC are of African, South Asian, East Asian, Finnish, Non-Finnish European, and Admixed American (Latino) ancestry[19]. About half of all functional-variants in drug-related genes ($M = 54\%$, $SD = 15.2\%$) are unique to only one of the six populations and only 0.1% of functional-variants occur with an AF $\geq$ 0.1% across all populations. Consequently, this results in drug-related genes that have a high risk of functional-variants depending on geographic ancestry.

For instance, using a cutoff of CAP>1%, we found that 231 drug-related genes have functional variants in the cohort of European ancestry compared to 298 genes with functional variants for the cohort of African ancestry.

Nevertheless, 114 drug-related genes showed a CAP score above 1% in each population indicating that there are genes with a similar world-wide pharmacogenetic relevance.

Not surprisingly, amongst those genes with the highest difference in CAP score between populations are many cytochrome P450s and phase II enzymes (Supplementary Table 5), as noted in previous studies of smaller population sizes[22]. Similarly, we observe drug target genes with markedly different CAP scores across populations. Among the target genes with the highest absolute CAP score difference are *VWF* (which is targeted by antihemophilic factor), *SIRT5* (targeted by suramin for treating sleeping sickness), and the gastric lipase *LIPF* (targeted by orlistat for obesity treatment). The latter has 65 functional-variants and the most frequent variants differ especially between African and East Asian cohorts (CAP 8% vs 51%). Target genes with high subpopulation differences also include several targets for antineoplastic agents, such as the olaparib-target *PARP1*, for which the CAP score ranges from 10.2% in patients of African ancestry to 69.6% in Latino patients. While the efficacy of olaparib depends on the tumor genome and not the germline, the risk to carry germline-originated variants in the tumor should not be ignored. We also observed population differences in the nucleoside transporter *SLC28A1*. While the CAP score is 4% in Non-Finish Europeans, individuals with an East Asian ancestry have a risk of 60%. Interestingly, several variants in *SLC28A1* have been associated with different outcomes in non-small cell lung cancer and breast cancer[42,43] when treated with gemcitabine, suggesting that variant differences across the populations may be involved.

**Analysis of the DRP score reveals a population-specific risk for several drugs**

Of the 1,236 FDA approved drugs considered, 241 have more than 10% absolute difference in DRP scores between at least two sub-population cohorts and 24 of these have more than 30% DRP difference (Supplementary Table 6). Out of this subset of drugs, 11 belong to the 100 most prescribed drugs in the US and 28 are recommended worldwide by the WHO for their therapeutic use, including oxcarbazepine, amobarbital and dolasetron. 312 of the 1,236 drugs have a high risk (DRP>1%) in all six sub-populations (Fig. 4A, and the DRP top 20 drugs stratified by population are illustrated in Fig. 4B).

Well-known differences, such as response to disulfiram (treatment for chronic alcoholism), are recapitulated in the data (Fig 4B). Specifically, the genetic variant E487K in the disulfiram target *ALDH2* (rs671) is seen in the ExAC East Asian population at similarly high frequencies as seen in previous genetic studies[44].

The different responses in the asthma-medication salbutamol and the blood-thinner warfarin have been attributed to variants in their respective drug targets, including R16G in *ADRB2* (rs1042713) for salbutamol[45] and 1639G>A (rs9923231) in *VKORC1* for warfarin[46]. Since the well-known response altering variants were not annotated by mutation prediction software as functional-variants, we did not expect to see the drugs appear high in our ranked list of risk differences across the populations (see discussion). Nevertheless, our analysis shows that salbutamol still has a high risk ratio between populations, caused by 29 variants with a dominant contribution from one variant separating the individuals of Finnish ancestry from African ancestry (rs201257377, N69S, $AF_{FIN}$=0.01). To our knowledge this variant has not been functionally characterized or previously associated with salbutamol response. Similarly, we observe 19 functional-variants in the warfarin target *VKORC1* that are population-specific, including a functional-variant observed most frequently in individuals of Non-Finnish European or Latino ancestry, (rs61742245, D36Y, $AF_{NFE}$=0.003, $AF_{Latino}$=0.001), that has been previously associated with predisposition for warfarin resistance[47]. However, 16 of the functional-variants may be novel risk factors including a functional-variant primarily observed in individuals of East Asian ancestry (R53S, ENST00000394975.2:c.157C>A, $AF_{EAS}$=0.001). Using a recent protein 3D model[48,49] of *VKORC1,* we mapped the R53S variant to the putative warfarin binding pocket (Fig. 3B). Furthermore, analysis of coevolution in the protein using EVfold[50] shows that R53 is strongly coupled to other residues in the protein and changes in this site are predicted by EVmutation[51] to affect protein fitness due to epistatic variant effects (Supplementary Fig. 5). Together, this suggests that this mutation might be negatively associated to warfarin binding.

Triflusal, a treatment for stroke re-occurrence, targets four genes (*PTGS1* (also known as Cox-1), *NOS2*, *NFKB1*, and *PDE10A*) that together have more functional-variants in the African population than in any other population ($DRP_{AFR}$=37%, Fig. 4B). This difference between populations is mainly due to a SNP in *NOS2,* which occurs in the population of African ancestry with higher than average frequency (rs3730017, $AF_{AFR}$=19% vs $AF_{global}$=4%) and while not functionally characterized, has been associated with protection against cerebral malaria[52]. In *PTGS1*, three functional-variants have allele frequencies above 0.1% in the cohort of African ancestry. The most frequent variant (rs5789, L237M, $AF_{AFR}$=0.5% vs $AF_{global}$=1.7%) lies on the dimer interface and has previously been associated with reduced metabolic activity of the enzyme[53]. A second variant is an indel, which is predicted to result in the total loss of protein function ($AF_{AFR}$=0.3% vs $AF_{global}$=0.02%). The effects of the third functional-variant common in the African cohort (rs139956360, E259A, $AF_{AFR}$=0.2% vs $AF_{global}$=0.02%) on enzyme activity or drug binding is less clear from the three-dimensional structure of the protein and would require further exploration. Since triflusal is prescribed for prophylactic use in the same way as aspirin for stroke prevention, it is clearly worth further investigating the effects of these observed functional-variants.

**Population differences in functional-variants for cancer drugs.**

Our results also highlight a large DRP variability of cancer drugs between the populations. While for many of these drugs not the germline but the tumor genome are relevant for drug action, germline DRPs of these drugs give an estimate of the population risk to possess potentially resistance-causing variants in the tumor and should be screened accordingly. For instance, the DRPs of taxanes (docetaxel, paclitaxel and cabazitaxel) are 30 percentage points higher in the cohorts of South Asian and European ancestry compared to the cohort of African ancestry (DRP$_{SAS/NFE}$=85% vs DRP$_{AFR}$=45%) due to functional-variants in the four taxane targets, *TUBB1*, *MAP2*, *MAP4* and *MAPT*. Among these are three distinct positions in *TUBB1* (Q43P/H, R307C, R359W) that occur with comparably high frequencies in the South-Asian population. While Q43P (AF$_{SAS}$=14%) has recently been associated with decreased progression-free survival in urothelial cell carcinoma when treated with cabazitaxel[54], less is known about the effects of the other two variants. Mapping the affected residues onto the three dimensional structure of docetaxel bound to tubulin (PDB ID: 1tub[55]) shows that R359 interacts with the drug (Fig. 3C). The effect of R307C is less obvious from structural observations as it does not lie very close to the binding site or the interface between the monomers in the polymer (R307 to K124 < 15 Å, mapped on PDB ID: 3j6g[56]).

## Discussion

In this study, we analyzed the extent of functional genetic variation in drug-related genes and its implication for 1236 FDA-approved drugs in exome sequencing data of 60,706 individuals. We show that not only the risk of carrying functional-variants in ADME-related genes, but also in drug targets is high for an individual patient. For ADME-genes this observation is in line with previous studies[12,15,18], but novel for drug-target genes. We observed functional-variants in 98% of the drug-related genes and at least one high confidence LoF variant in 93% of the genes. The prevalence of functional-variants in drug-related genes is thus higher than previously shown[18]. When considering drug target genes for the 100 most prescribed medications in the US the probability of carrying at least one functional-variant is above 80% for each patient. Together with the high risk for clinically actionable variants in ADME genes (98%[12]) these findings indicate that genetic variability may contribute significantly to observed differences in drug response between patients.

While individualized cancer therapies often focus on the somatic variants present only in tumor tissue, we can show that functional germline variants, which are routinely masked out in the analysis of somatic variants, are common in many cancer drug targets. By excluding germline variants that the tumor inherited from its progenitor cell from cancer genome analysis in the context of therapeutic decision-making may thus result in the oversight of important determinants for treatment response or resistance development. To what extent the tumor genome varies from the germline genome, is dependent on patient and cancer type. Loss of heterozygosity, where the germline allele is lost in the disease progression and copy number alterations can indeed result in drastic changes between genetic variants observed in the normal tissue of a patient and the cancer[30,57]. The high prevalence of variants in systemic cancer therapy targets, such as *KDR* for sorafenib, further indicates, that the germline variants of target genes in addition to ADME genes should be considered for clinical decision making.

Geographic ancestry is a well-established confounding factor for drug response, but few drugs have been assessed in their efficacy across global populations. Even where clinical trials have been carried out in different populations, particularly non-European and non-Asian individuals remain understudied. By calculating risk probabilities for drugs and different populations, we showed that the frequency of functional-variants in drug-related genes varies widely across populations. Even for drugs where population differences in response are observed, additional patient groups may be at high risk of altered PD due to genetic variants in drug targets. Especially for drugs commonly used around the world, such as those on the WHO Essential

Medicines list, this could result in large numbers of patients with reduced drug efficacy in some, but not all, of the populations they are applied in.

The analysis in this study relied on external data for drug variant annotation and drug-gene associations. Even though it was possible to estimate the burden of functional variation in drug-related genes and quantify to which extent individual drugs may be affected, there remain certain limitations. First of all, even manually curated drug-target associations and pharmacogenomics data are susceptible to spurious annotations. For example, some subunits of the GABA receptors including *GABRA4* are generally thought to give rise to receptors resistant to classic benzodiazepines such as diazepam[58], but have been annotated as targets for some benzodiazepines. Comparison to a different, independently curated set of drug-target associations[59] further shows that annotation of drug – target pairs does not always agree. Furthermore, to quantify the real risk for a drug, drug-specific ADME-gene relations should be incorporated into the DRP calculation. For example, optimal warfarin dosing is known to be dependent on variants in *CYP2C9* in addition to *VKORC1*[60] and variants in the ADME-gene *UGT1A1* are documented to contribute to different responses to the cancer drug irinotecan around the globe[61]. Unfortunately, comprehensive inclusion of ADME-genes in the DRP calculations is currently not possible because sufficient data for ADME-genes is lacking for most FDA approved drugs including the relative contribution of each enzyme. Our DRP estimates thus probably still underestimate the drug-specific risk of functional variation as well as population differences.

The vast majority of variants in drug-related genes considered in this study has not been seen previously and thus lacks validated knowledge about their functional impact on drug efficacy. We therefore had to rely on predictions of their impact on protein function. The probabilities presented are based on the assumption that the functional classification is correct and represents enzyme activity or drug efficacy. The relative risk between genes is based on the assumption that there has not been a significant bias in assessment when genes already have known deleterious mutations. That these assumptions are not always correct, follows from the fact that variant classification tools are not exact, are often trained on disease-causing variant sets only, have issues with circularity in the classifier training data, and fail to sub-classify mutations[62]. Especially the distinction of activating and deactivating effects could be crucial for the downstream effects on therapy.

This discrepancy between observed and predicted functional-effects can be illustrated on the well-studied PGx variants in the anti-asthmatics target *ADRB2* (R16G/rs1042713, Q27E/rs1042714 and T164I/rs1800888) that all are classified as benign[45,63]. To alleviate this problem, one could include additional prediction algorithms, which comes at the risk of reduced specificity (in some cases more than half of all non-synonymous variants were classified as functional[15]) as all currently available methods have their individual drawbacks[64]. Reliable computational classification methods for variant effects on drug response remain scarce due to insufficient training data [64], but may arise in the future if efforts are increased to create such data, for example using novel high throughput methods such as deep mutational scans[65,66]. For the present study we chose a conservative approach to variant annotation that requires the complete loss of the protein product – which should have a marked impact on the drug – or the consensus prediction of two independent prediction tools at the expense of missing some known variants (Fig. 3A). It is thus not unlikely that the effect of the functional-variants is still underestimated in our study.

**Sequencing data.** The use of whole exome sequencing data comes with the intrinsic limitation that only variants in protein coding regions can be detected, potentially missing pharmacologically relevant non-coding variants[67] or larger structural changes of the genome. Furthermore, even at low false-positive rates many called variants can be inaccurate[68] and several pharmacologically relevant gene families – namely CYPs, HLA and UGTs – are at high risk for variant calling errors due to the complex genetic structure of their loci[69,70]. While members of the cytochrome P450 family have indeed been found to be problematic in short-read sequencing[22], this does not apply for most other drug-related genes[15,18]. To reduce the false-positive variant calls in our survey, we included only variants of sufficient locus coverage and high quality.

Furthermore, the ExAC cohort is very large in total, but not all populations are represented equally[19]. The power to detect very rare variants thus differs by an order of magnitude between the individual populations (from 0.01% AF for the Finnish and East Asian populations to 0.001% for Non-Finnish European). Due to legal restrictions in the underlying exome sequencing projects, sample-specific data including haplotype phase is missing also in ExAC. Epistatic effects of variants could thus not be investigated, even though they are known to exist. For example, while the single variant rs12248560 (CYP2C18*17) results in increased *CYP2C19* activity, the combination with another variant (rs28399504) is associated with loss-of-function of the protein (CYP2C19*4B)[15].

**Implications**. Many major medical institutions have started implementing genotyping protocols for preemptive pharmacogenetic testing[71-73]. However, these usually focus on a small number of ADME-genes[12] and often only test a subset of established actionable variants using microarrays[74]. While these arrays facilitate fast and cheap screening, we show here that the vast majority of variants in drug-related genes seen in the human population is not covered. We further want to motivate that the number of genes with pharmacogenomic variants should systematically include genes implicated in drug mechanism even though only very few examples in such genes have yet been characterized well enough to be part of a dosing guideline. Furthermore, with allele frequencies below 0.1%, many functional-variants in drug-related genes are so rare that they cannot be observed in clinical trial cohorts, but may contribute to adverse events or diffuse lack of efficacy post-marketing. In the future, this should be in all phases of clinical drug development and the effects of genetic variants in genes associated with PD and PK of the drug candidate should be systematically characterized.

In conclusion, large-scale sequencing efforts can be used to identify and quantify the extent of genetic variation in genes relevant for drug action and metabolism. Identification of such variants is only the first step towards better treatment decisions. Newly identified variants of pharmacogenomics importance require validation and ultimately updated dosing guidelines. The development of quality-controlled and patient-centered software solutions to combine available knowledge of pharmacologically actionable variants with a patient's genome as well as fast and accurate approaches (experimental and computational) to functionally classify novel variants will thus be of high importance for a future of personalized medicine.

## Materials and Methods

**Data selection and handling**

Known pharmacogenomics associations between drugs and genetic variants were retrieved from PharmGKB[27]. Data about drugs and drug-related genes was collated from DrugBank 5[21].

Information about drug approval status, ATC code, and details about the drug – gene relationship (target, pharmacological action and action type) were extracted from the xml file using python. We further obtained a list of the top 100 most prescribed drugs of 2013 from drugs.com[34] and the list of WHO essential medicines by parsing the Index of the 19th WHO Model List of Essential Medicines[33]. Drugs obtained from the top 100 list and WHO essential medicines catalog were mapped to DrugBank compounds and those where this was not possible were excluded. Relations between hyaluronic acid and human gene targets as well as between dihydropyridines and skeletal *CACNA1S* were removed because the literature in the database entry did not support the pharmacological involvement of these pairs. We further removed Ethanol from the list of WHO essential medicines because it is listed as a surface disinfectant and thus not dependent on the patient's cellular targets.

Drug target genes were extracted from the drug – gene relationships in DrugBank, by filtering this set for only those relations with established pharmacological action flag and in which the gene is annotated as drug target. Based on previous studies a list of pharmacologically relevant cellular receptors, metabolic enzymes and nuclear receptors was obtained from to recent pharmacogenomics surveys[15,22] and comprises the set of ADME-genes.

Genetic variant information including variant types, allele frequencies and deleterious prediction scores were extracted from the ExAC VCF file (release 0.3) downloaded from the ExAC FTP server[19]. Multi-allelic variants were split using vcflib breakmulti (https://github.com/vcflib/vcflib) and synonymous variants were excluded. We then calculated for each variant the allele frequency (AF) in the full cohort as well as in each ExAC population separately by dividing the allele count (AC) by the allele number (AN). Following information about ancestry were used: AFR=African, SAS=South-Asian, EAS=East-Asian, FIN=Finnish,

NFE=Non-Finnish Eurpean, AMR=Admixed American/Latino. We further excluded variants whose loci were not observed at least once in every geographic population and in 50% of all possible samples (i.e., minimal allele number of 60,706). After adding unique IDs to the variants based on chromosome position, reference and alternative gene, we removed duplicates.

Identifier mapping, filtering and annotation was performed using the Konstanz Information Miner (KNIME) workflow system[75] and the Python programming language (Python Software Foundation, https://www.python.org/).

**Variant subsets**

To evaluate variants with functional effects in the ExAC catalog, we created a subsets of variants with functional effects ("functional-variants"): 1) loss-of-function variants affecting stop codons, splice sites and shifts in the reading frame as annotated by the Loss-Of-Function Transcript Effect Estimator (LOFTEE) tool[76] in the ExAC VCF file, and 2) variants predicted to have a damaging effect on the protein as predicted unanimously by PolyPhen-2 [23] ('possibly damaging' or 'probably damaging') and SIFT[24] ('deleterious') as annotated in the ExAC VCF file. Functional-variants with allele frequencies above 0.5 were excluded from this set after observing that there are annotation or reference genome mapping problems. For each gene we calculated the fraction of common (AF >= 0.1%) and rare (AF < 0.1%) alleles.

**Computation of cumulative probabilities for drugs and their related genes**

To quantify the risk of an individual person in the population to carry functional-variants in a particular gene, we define the "cumulative allele probability" (CAP) statistic, which captures both the number of functional-variants and their allele frequencies per gene. Formally, this score is the probability for an individual to carry at least one variant allele a of the observed alleles A in a gene g.

$$CAP(g) = 1 - \prod_{a \in A}(1 - AF(a))^2$$

Two types of CAP scores were calculated, one for all functional-variants in a drug-related gene and one based only on LoF variants.

To estimate how much each drug can be affected by functional-variants in its target genes, we further define the drug-specific "drug risk probability" (DRP) score by combining the CAP scores for all drug target genes. Formally, the DRP score is defined as

$$DRP(D) = 1 - \prod_{g \in G} \prod_{a \in A_g}(1 - AF(a))^2$$

Here G is the set of all target genes for drug D, as documented in DrugBank, and $A_g$ the set of all variant alleles observed in gene g.

Correlation analysis of the DRP scores with the number of targets was performed using linear regression with ordinary least squares fitting using the Python package statsmodels[77] to compute the coefficient of determination $r^2$.

**Statistical Analysis of population differences**

Population comparisons for CAP and DRP scores were performed using the absolute risk difference (RD) metric.

$$RD = |P(\text{event in group 2}) - P(\text{event in group 1})|$$

The RD for a drug was calculated by subtracting the score from population with the smallest DRP score from the score of the population with the highest DRP. To identify for which drugs a population has above or below average risks (Fig. 4b), we further calculated all pairwise risk differences between populations from which we then computed the population-specific mean RDs.

**Detailed variant analyses in case studies**

Protein structures for the porcine *TUBB1* homologue (PDB IDs: 1tub[55], 3j6g[56]), *ADRB2* (PDB ID: 2rh1[78]), *PTGS1* (PDB ID: 3n8w[79]) and *NOS2* (PDB ID: 4nos[80]), were obtained from the Protein Data Bank. Recently published homology models for *VKORC1* were downloaded from the supplement of the respective publications[48,49]. Co-evolution analysis of residues was done using plmc-based EVcouplings[50] and based on jackhmmer[81] alignments created with the Uniprot entries of the respective protein as queries against the Uniref100 database[82] (release 01/2017). Alignment columns with more than 70% gaps and sequences with more than 50% gaps were excluded from the model. Functional impact was predicted using EVmutation[51] and, in the case of *VKORC1*, compared to experimental warfarin binding data[49]. Protein structures were analyzed and rendered using the UCSF Chimera package from the Computer Graphics Laboratory, University of California, San Francisco[83].

**Statistical analysis and code availability**

Statistical analysis of the data set was performed in jupyter/IPython notebooks[84] using pandas[85] and other packages of the SciPy stack[86]. The code used to analyze the data set and produce the figures will be made available on github.

**Acknowledgements:**

We would like to thank Ruomu Jiang for initial help with handling genetic variation data sets, Benjamin Schubert, Fabian Aichler, and Ulrich Mansmann for helpful discussions about the statistical analysis performed in the paper and Thomas Hopf for support in using the EVmutation toolbox.

**Author's contributions:**

CPS, DSM and OK designed the study, CPS analyzed the data, DSM and OK helped analyzing the data, RT and MS provided expertise of pharmacogenetics and genomics and contributed in interpretation of the data, CPS and DSM wrote the manuscript, all authors contributed to editing the manuscript.

**Funding:** This work was also supported in part by the Robert Bosch Foundation, Stuttgart, Germany and the European Commission Horizon 2020 UPGx grant (668353).

*The authors declare no conflict of interest*.

**Abbreviations**

AF             allele frequency

ADME           absorption, distribution, metabolism and excretion

| | |
|---|---|
| ExAC | Exome Aggregation Consortium |
| PD | pharmacodynamics |
| PK | pharmacokinetics |
| GWAS | genome-wide association study |
| LoF | loss-of-function |
| RMSE | root mean square error |
| CAP | cumulative allele probability |
| DRP | drug risk probability |
| WHO | World Health Organisation |

Figure 1

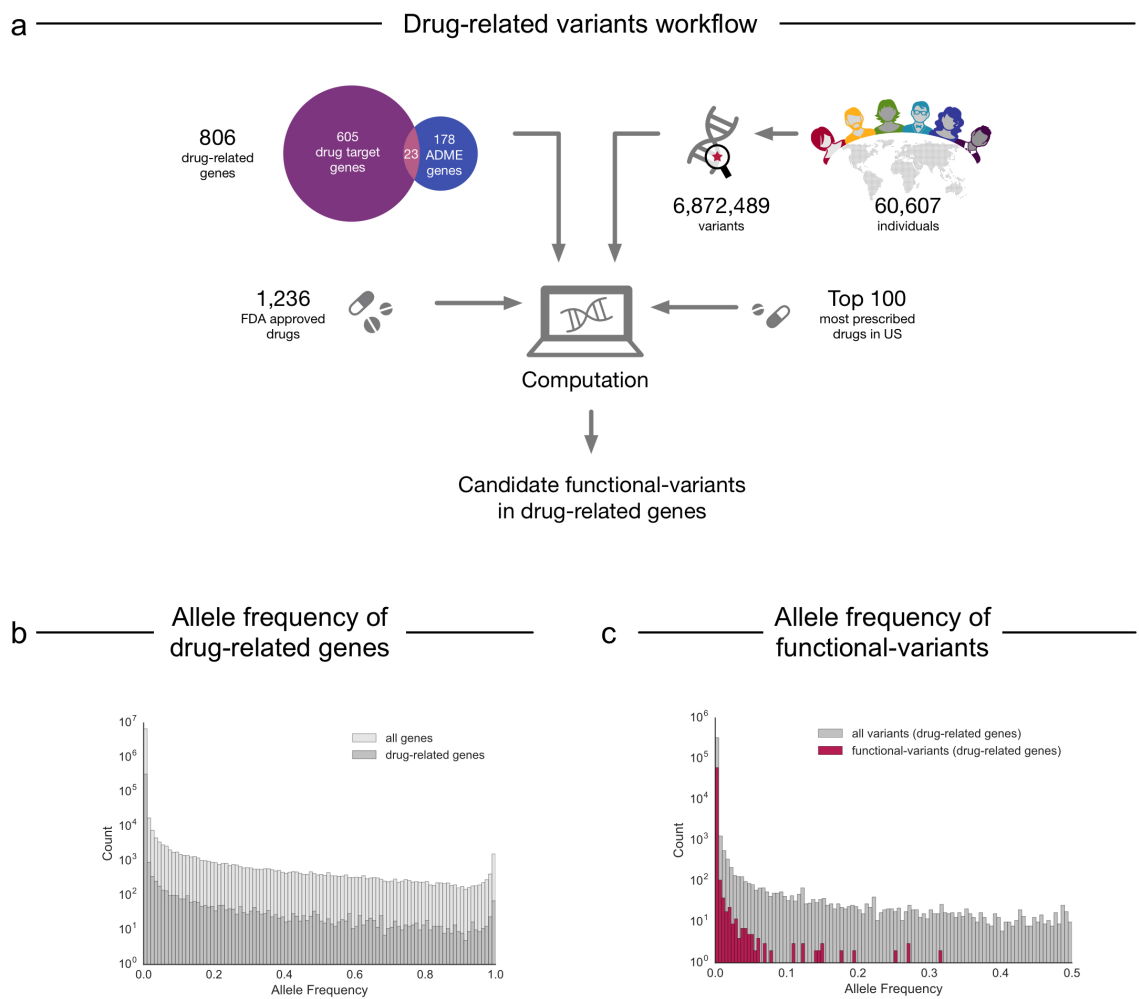

**Figure 1. Analysis of genetic variation in drug-related genes**. a) The analysis pipeline consisted of collation of exome data from ExAC[19], identification of drug – gene relationships from DrugBank[21] and prescription information[34] followed by filtering steps and subsequent computational analysis to investigate drug-specific risks of pharmacogenetic alterations in patients. b) Comparison of the allele frequency distribution between non-synonymous variants of all human genes (n=17,758) and non-synonymous variants in drug-related genes (n=806) collated from ExAC. c) Comparison of the allele frequency distribution between functional-variants as predicted by LOFTEE[76], Polyphen-2[23] and SIFT[24] and all non-synonymous variants in the drug-related genes.

Figure 2

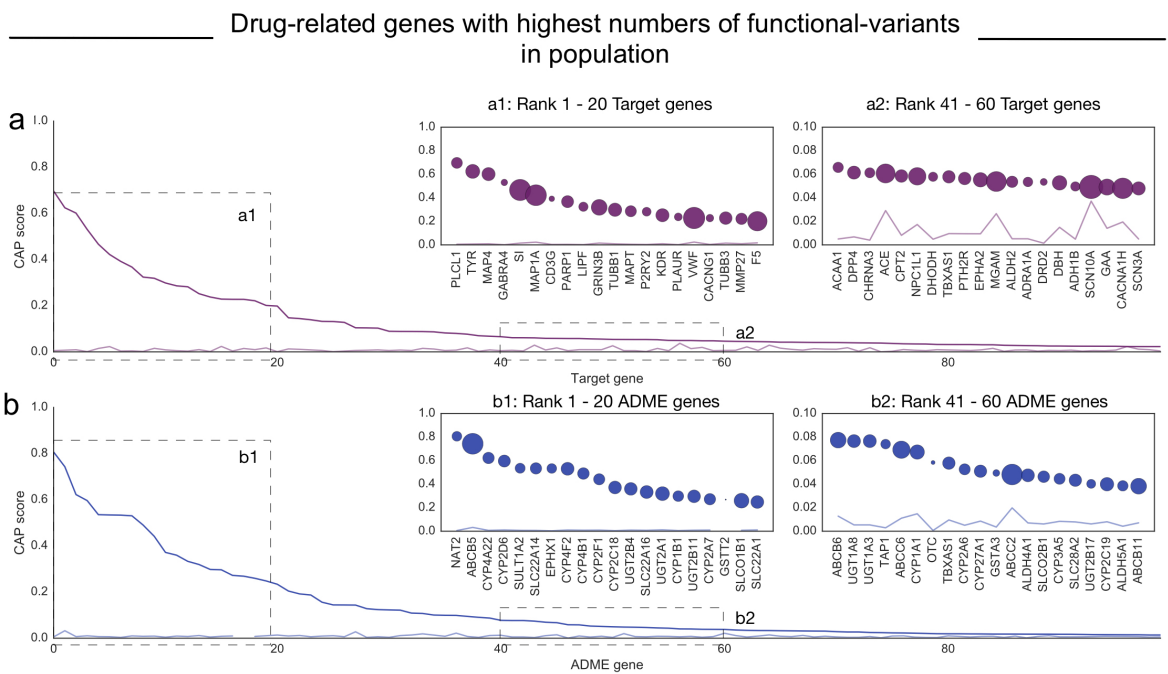

**Figure 2. Drug-related genes with highest probability of having functional-variants.** a) Protein-centered cumulative allele probability (CAP) scores for the 100 drug targets with highest scores (purple) and the contribution of CAP scores as determined from rare variants alone (light purple). a1) Top 20 target genes with highest CAP score, a2) Examples of target genes with lower CAP scores, b) 100 ADME-genes with highest CAP scores (blue), and the corresponding CAP score determined from rare variants alone (light blue). b1) Top 20 ADME-genes with highest CAP scores, b2) Examples of ADME-genes with lower CAP scores. Bubble size corresponds to the number of functional-variants observed for the respective gene.

Figure 3

a ──── Functional-variants in targets of 1236 FDA approved drugs ────

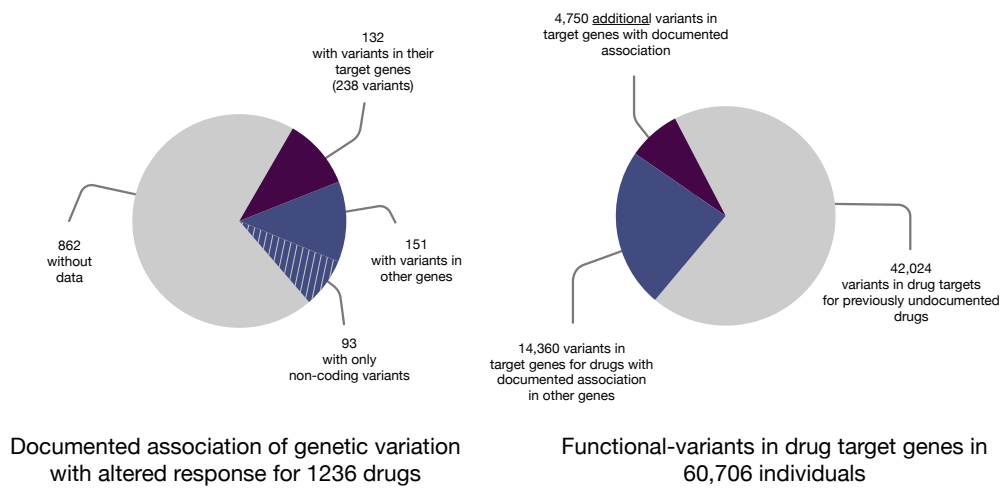

Documented association of genetic variation
with altered response for 1236 drugs

Functional-variants in drug target genes in
60,706 individuals

b ──── Warfarin target: VKORC1 ────   c ──── Docetaxel target: TUBB1 ────

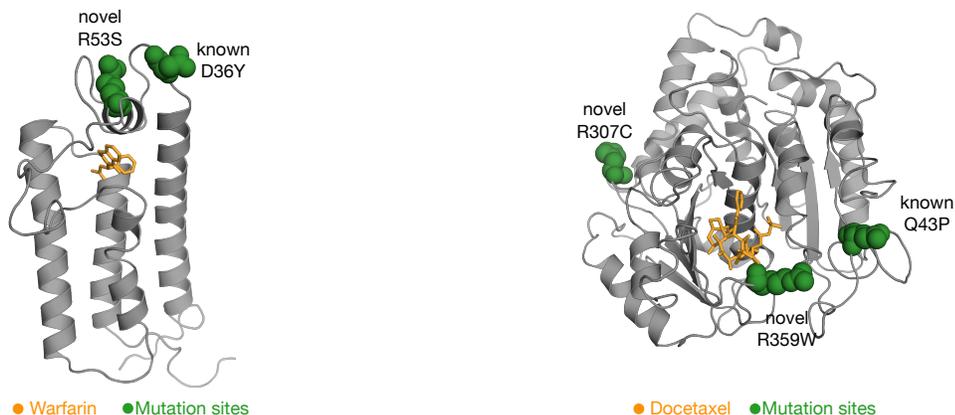

● Warfarin  ● Mutation sites        ● Docetaxel  ● Mutation sites

**Figure 3**. **Knowledge gap between observed genetic variants in the population and documented pharmacogenomics data**. a) Availability of documented pharmacogenetic associations for 1,236 FDA-approved drugs in public repositories such as the PharmGKB database[27] (left), is less abundant than functional-variants observed in the population for the drug target genes (right). b) and c) Examples of known and novel genetic variants (green) in the target genes of warfarin and taxanes that could affect drug efficacy due to effects on the binding site (ligand highlighted in purple).

Figure 4

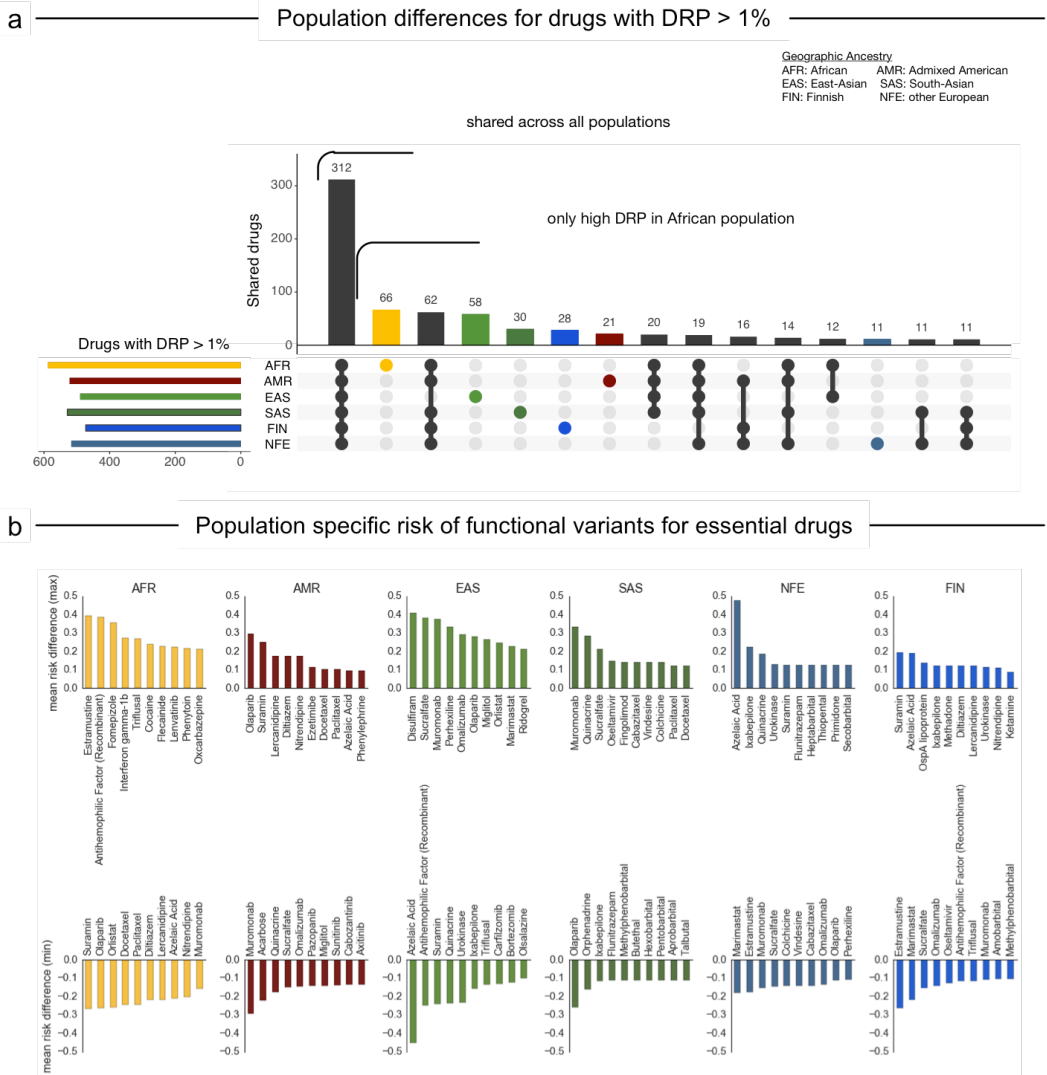

**Figure 4. Variability of drug risk probabilities across populations.** a) Number of drugs with shared (black) or private (colored) drug risk probabilities (DRP) for functional-variants in their pharmacological target genes greater than 1%. DRP scores were calculated by aggregating the risk of functional variation across all documented pharmacological target genes of that drug. b) Drugs with highest (top) or lowest (bottom) mean DRP difference compared to all other populations indicating for which this population is at higher/lower risk of encountering functional-variation in the target for a drug and thus higher/lower impact on drug effect.